\input harvmac
\input epsf
 \def \trace{ \mathop{ \rm
trace}\nolimits} 
\Title{DFTUZ /97-10}
{\vbox{\centerline{Excitations and S-matrix for $su(3)$ spin chain }
\vskip2pt\centerline{combining $\{3\}$ and $\{3^{*}\}$
representations}}} \centerline{J. Abad and M. R\'{\i}os}
\centerline{Departamento de F\'{\i}sica Te\'{o}rica, Facultad de
Ciencias,} \centerline{Universidad de Zaragoza, 50009 Zaragoza,
Spain} \bigskip
\bigskip
\vskip .3in
\centerline{ \tenbf Abstract}

The associated Hamiltonian for a $su(3)$ spin chain combining $\{3\}$
and $\{3^{*}\}$ representations is calculated. The
ansatz equations for this chain are obtained and solved in the
thermodynamic limit, and the ground state and excitations are
described. Thus, relations between the number of roots and the number
of holes in each level have been found . The excited states are
characterized by means of these quantum numbers. Finally, the exact
$S$ matrix for a state with two holes is found. \bigskip

\bigskip
\bigskip
\Date{}

\vfill
\eject

\newsec{Introduction}

The Yang-Baxter equation (YBE)
\ref\ri{C.N. Yang, Phys. Rev. Lett. 19, (1967) 1312.} \ref\rii{R.J.
Baxter, Ann. Phys. (N.Y.) 70, (1972) 193.} and the quantum inverse
scattering method (QISM)
\ref\riii{E.K. Sklyanin, L.A. Takhtadzhyan and L.D. Faddeev, Teor.
Mat. Fiz. 40, (1979) 194.}
have contributed to find and solve a lot of many body quantum systems.
The best known system is the Heisemberg model, which was solved by
Bethe \ref\riv{H. Bethe, Z. Phys. 71, (1931) 205.}. This model can be
derived from the YBE using the $su(2)$ Lie algebra. Generalizations
of this model have been obtained using other Lie algebras
\ref\rv{H.J. de Vega, Int. J. Mod. Phys. A 4, (1989) 2371.}-%
\nref\rvi{P.P. Kulish and N.Y. Reshetikhin, Sov. Phys. JETP 53
(1981).}%
\ref\rvii{J. Abad and M. R\'{\i}os, Univ. de Zaragoza preprint DFTUZ
94-11 (1994), cond-mat 9609220.}.

An interesting problem is to derive integrable models where the chain
is formed by two kind of states. The original work, an alternating
chain with $s=1/2$ and $s=1$, was presented in Ref. \ref\rviii{H.J.
de Vega and F. Woynarovich, J. Phys. A 25, (1992) 4499.}. Later,
several works,
using several Lie algebras, have been studied \ref\rix{J. Abad and M.
R\'{\i}os, Phys. Rev. B 53, (1995) 14000, J. Phys. A 29, (1996) L1.}.
In this systems it is possible to solve the ansatz equations in the
thermodynamic limit \ref\rxxiv{O. Babelon, H.J. de Vega and C.M.
Viallet, Nucl. Phys. B 220, (1983) 13.}.
This allows us to describe the system by means of the quantum numbers
and so we would be able to one can find the ground state and the
excited estates \ref\rx{C. G\'omez, M. Ruiz-Altaba and G. Sierra,{\it
Quamtum Groups in Two-Dimensional Physics}, Cambridge Univ. Press,
1996.}-%
\nref\rxxi{L. Faddeev and L.A. Takhtajan, J. Sov. Math. 24,
(1984) 241.}%
\ref\rxi{L. Fadeev, {\it Lectures in Quantum Inverse Scattering
Method}, in Nankay Lectures on mathematical Physics. Edited by Song
Xing-Chang.World Scientific, Singapore, 1987.}. Besides, the $S$
matrix for the scattering of excitations can be determined
\ref\rxii{L. Mezincescu and R.I. Nepomechie, {\it Exact $S$ matrices
for integrable quantum spin chains}, UMTG-180 1994.}- \nref\rxxii{P.
Fendley and H. Saleur, Nucl. Phys. B 428, (1994) 681.}%
\nref\rxxiii{R. Shankar and E. Witten, Phys. Rev. D 17, (1978)
2134.}%
\nref\rxiii{H.J. de Vega, L. Mezincescu and R.I. Nepomechie,
J. Mod. Phys. B 8, (1994) 3473.}%
\nref\rxiv{H.J. de Vega, L. Mezincescu and R.I. Nepomechie, Phys.
Rev. B 49, (1994) 13223.}%
\nref\rxv{S.R. Aladim and M.J.
Martins, J. Phys. A 26, (1993) L529.}%
\nref\rxvi{S.R. Aladim and M.J.
Martins, J. Phys. A 26, (1993) 7287.}%
\ref\rxvii{M.J. Martins, J. Phys. A 26, (1993) 7301.}.

In this paper we use the $su(3)$ rational solutions of the YBE and we
form a chain combining $\{3\}$ and $\{3^{*}\}$ representations. For
the alternating chain we find the Hamiltonian, which contains a
coupling of three neighboring site pieces. We solve the ansatz
equations and we deduce the root and hole densities. The relations
between these densities allow us to describe the ground and excited
states. In the last section we calculate the exact $S$ matrix for the
two-hole scattering.

\newsec{The model and the Hamiltonian}

In this section, we are going to construct an alternating chain that
mixes the $\{3\}$ and $\{3^{*}\}$ representations of $su(3)$. We use
the rational solutions of Yang-Baxter equation. If we take the
$\{3\}$ representation as auxiliary space and $\{3\}$ as site space,
we have the operator \eqnn\eai
$$\eqalignno{L^{(\{3\},\{3\})}(u)
&=(1-iu) \sum_{j=1}^{3} e_{j,j} \otimes e_{j,j} -iu \sum_{j,k=1 \atop
j \neq k}^{3} e_{j,j} \otimes e_{k,k}
+ \sum_{j,k=1 \atop j \neq k}^{3} e_{j,k} \otimes e_{k,j} . & \eai
\cr }
$$

For the $\{3\}$ as auxiliary and $\{3^{*}\}$ as site, the operator is
\eqnn\eaii
$$\eqalignno{L^{(\{3\},\{3^{*}\})}(u)
&=({1 \over 2}-iu) \sum_{j=1}^{3} e_{j,j} \otimes e_{j,j} -({3 \over
2}-iu) \sum_{j,k=1 \atop
j \neq k}^{3}
e_{j,j} \otimes e_{k,k}
- \sum_{j,k=1 \atop j \neq k}^{3} e_{j,k} \otimes e_{j,k}, \qquad
&\eaii \cr }
$$
with $(e_{l,m})_{i,j}=\delta_{l,i}\delta_{m,j}$.

We consider a chain with N sites (N even) in which the site spaces
are alternating in the representations $\{3\}$ and $\{3^{*}\}$. The
monodromy matrix, which describes the transportation along the chain,
is defined by %
\eqn\eaiii{ T_{a,b}(u,\alpha) = L_{a,a_{1}}^{(\{3\},\{3\})}(u)
L_{a_{1},a_{2}}^{(\{3\},\{3^{*}\})}(u+\alpha) \ldots
L_{a_{N-2},a_{N-1}}^{(\{3\},\{3\})}(u)
L_{a_{N-1},b}^{(\{3\},\{3^{*}\})}(u+\alpha) , }
where the indices are in the auxiliary space and $\alpha$ is an
arbitrary parameter.

Since $L^{(\{3\},\{3\})}(u)$ and $L^{(\{3\},\{3^{*}\})}(u)$ operators
verify the YBE, then $T(u,\alpha)$ also verifies it %
\eqn\eaiv{ R(u-v) \cdot (T(u,\alpha) \otimes T(v,\alpha)) =
(T(v,\alpha) \otimes T(u,\alpha)) \cdot R(u-v) ,
}
with
\eqnn\eav
$$\eqalignno{R(u)
&=(1-iu) \sum_{j=1}^{3} e_{j,j} \otimes e_{j,j} -iu \sum_{j,k=1 \atop
j \neq k}^{3} e_{j,k} \otimes e_{k,j}
+ \sum_{j,k=1 \atop j \neq k}^{3} e_{j,j} \otimes e_{k,k} . & \eav
\cr }
$$

Following the standard procedure, we take the transfer matrix as the
trace, on the auxiliary space, of the monodromy matrix %
\eqn\eavi{F(u,\alpha) = \trace [T(u,\alpha)] . }
Due to YBE, the transfer matrices commute for different values of the
argument
\eqn\eavii{\left[F(u,\alpha),F(v,\alpha)\right]=0. }

The Hamiltonian of that system is defined by the first derivative of
the transfer matrix,
\eqn\eaviii{H(\alpha) = \left. {d \over du} \ln (F(u,\alpha)
\right|_{u = 0}.
}
Collecting the diverse terms, the Hamiltonian becomes %
\eqn\eaix{H(\alpha) = {i \over \bar{\rho}(\alpha)} \sum_{j=1 \atop j
\, odd}^{N-1} 	h_{j,j+1}^{[1]} + {i \over c_{1}\bar{\rho}(\alpha)}
\sum_{j=1 \atop j \, 	odd}^{N-1} h_{j,j+1,j+2}^{[2]} \, ,
}
with
\eqna\eax
$$\eqalignno{
& \left( h_{j,j+1}^{[1]} \right)_{a,b;\beta,\gamma}
= [\dot{L}_{a,c}^{(\{3\},\{3^{*}\})}(\alpha)]_{\beta,\delta}
[L_{\delta,\gamma}^{(\{3^{*}\},\{3\})}(-\alpha)]_{c,b} & \eax a\cr
& \left( h_{j,j+1,j+2}^{[2]} \right)_{a,b;\beta,\gamma;c,d} =
[L_{a,e}^{(\{3\},\{3^{*}\})}(\alpha)]_{\beta,\delta}
[\dot{L}_{e,d}^{(\{3\},\{3\})}(0)]_{c,f}
[L_{\delta,\gamma}^{(\{3^{*}\},\{3\})}(-\alpha)]_{f,b} \qquad , &
\eax b\cr }
$$
and
\eqna\eaxi
$$\eqalignno{
& R(0) = c_{1}I & \eaxi a\cr
& [L_{a,b}^{(\{3\},\{3^{*}\})}(u)]_{\alpha,\beta}
[L_{\beta,\gamma}^{(\{3^{*}\},\{3\})}(-u)]_{b,c} = 	 \bar{\rho}(u)
\delta_{a,c} \delta_{\alpha,\gamma}\,. & \eaxi b\cr }
$$
Thus, we find
\eqnn\eaxii
$$\eqalignno{H(\alpha)=
& {2 \over {9+4\alpha^{2}}} \left\{ \sum_{i=1 \atop i \, odd}^{N-1}
\sum_{a=1}^{8} \lambda_{i}^{a} \otimes \bar{\lambda}_{i+1}^{a} +
\sum_{i=2 \atop i \, even}^{N}
\sum_{a=1}^{8} \bar{\lambda}_{i}^{a} \otimes \lambda_{i+1}^{a}
\right. & \cr
& + \sum_{i=1 \atop i \, odd}^{N-1}
\sum_{a,b,c=1}^{8} ({3 \over 2}d_{a,b,c}-\alpha f_{a,b,c})
\lambda_{i}^{a} \otimes \bar{\lambda}_{i+1}^{b} \otimes
\lambda_{i+2}^{c} & \cr
& \left. + {{5+4\alpha^{2}} \over 4} \sum_{i=1 \atop i \, odd}^{N-1}
\sum_{a=1}^{8} \lambda_{i}^{a} \otimes I \otimes
\lambda_{i+2}^{a} \right\} + {{41+12\alpha(\alpha+i)} \over
{9(9+4\alpha^{2})}} I , & \eaxii \cr
}
$$
where we have used the Gell-Mann matrices $\lambda$ and
$\bar{\lambda}$ for the $\{3\}$ and $\{3^{*}\}$ representations
respectively, being $d_{a,b,c}$ and $f_{a,b,c}$ the structure
constants of $SU(3)$. When $\alpha=0$, we obtain the simplest case.

\newsec{Diagonalization and ansatz equations} %
We have solved the chain that mixes the $\{3\}$ and $\{3^{*}\}$
representation of $su(3)$, using the method given in \rix. The
eigenvalue of the transfer matrix is
\eqnn\ebi
$$\eqalignno{\Lambda(u) =
& [a(u)]^{N_{3}}
	 [\bar{b}(u)]^{N_{3}^{*}}
	 \prod_{j=1}^{r} g(\mu_{j}-u) + [b(u)]^{N_{3}} 	 \prod_{i=1}^{r}
g(u-\mu_{i}) \times & \cr & \left\{
	 [\bar{b}(u)]^{N_{3}^{*}} \prod_{l=1}^{s} g(\lambda_{l}-u) +
[\bar{a}(u)]^{N_{3}^{*}} \prod_{k=1}^{s} g(u-\lambda_{k})
\prod_{n=1}^{r} {1 \over g(u-\mu_{n})} 	 \right\} , \qquad & \ebi \cr
}
$$
and the coupled Bethe equations are
\eqna\ebii
$$\eqalignno{
& [g(\mu_{k})]^{N_{3}} = \prod_{j=1 \atop j \neq k}^{r}
{g(\mu_{k}-\mu_{j}) \over g(\mu_{j}-\mu_{k})} 	 \prod_{i=1}^{s}
g(\lambda_{i}-\mu_{k}) &\ebii a\cr
& [\bar{g}(\lambda_{l})]^{N_{3}^{*}} =
	 \prod_{j=1}^{r} g(\lambda_{l}-\mu_{j}) 	 \prod_{i=1 \atop i
\neq
l}^{s}
	 {g(\lambda_{i}-\lambda_{l}) \over g(\lambda_{l}-\lambda_{i})} &
\ebii b\cr & \qquad k=1, \ldots ,r \quad ; \quad l=1, \ldots ,s \, ,
& \cr }
$$
with
\eqna\ebiii
$$\eqalignno{
& a(u)=1-iu &\ebiii a\cr
& b(u)=-iu & \ebiii b\cr
& \bar{a}(u)={1 \over 2}-iu &\ebiii c\cr & \bar{b}(u)={3 \over
2}-iu&\ebiii d\cr
& g(u)={a(u) \over b(u)} &\ebiii e\cr
& \bar{g}(u)={\bar{a}(u) \over \bar{b}(u)}. &\ebiii f\cr }
$$
It is convenient to set the parameterization \eqna\ebiv
$$\eqalignno{
& \mu_{j}=v_{j}^{(1)}- {i \over 2} &\ebiv a\cr
& \lambda_{j}=v_{j}^{(2)}- i. & \ebiv b\cr }
$$
Using such parameterization, the Bethe equations can be written
\eqna\ebv
$$\eqalignno{
& \left[{{v_{k}^{(1)}-{i \over 2}} \over {v_{k}^{(1)}+{i \over
2}}}\right]^{N_{3}} =- \prod_{j=1}^{r} {{v_{k}^{(1)}-v_{j}^{(1)}-i}
\over {v_{k}^{(1)}-v_{j}^{(1)}+i}} \prod_{l=1}^{s}
{{v_{l}^{(2)}-v_{k}^{(1)}-{i \over 2}}
\over {v_{l}^{(2)}-v_{k}^{(1)}+{i \over 2}}} &\ebv a\cr
& \left[{{v_{k}^{(2)}+{i \over 2}} \over {v_{k}^{(2)}-{i \over
2}}}\right]^{N_{3}^{*}} =- \prod_{j=1}^{r}
{{v_{k}^{(2)}-v_{j}^{(1)}-{i \over 2}} \over
{v_{k}^{(2)}-v_{j}^{(1)}+{i \over 2}}} \prod_{l=1}^{s}
{{v_{l}^{(2)}-v_{k}^{(2)}-i}
\over {v_{l}^{(2)}-v_{k}^{(2)}+i}} . & \ebv b\cr }
$$
We define the function
\eqn\ebvi{\phi(x)= \ln {{1+ix} \over {1-ix}} \equiv 2i\arctan x , }
and taking logarithms in \ebv{a,b}\ we obtain \eqna\ebvii{
$$\eqalignno{
&N_3 \phi(2v_k^{(1)})-
\sum_{j=1}^{r}{\phi(v_k^{(1)}-v_j^{(1)})}
+\sum_{l=1}^{s}{\phi(2v_k^{(1)}-2v_l^{(2)})} =2 \pi I_k^{(1)}, \
1\leq k \leq r, &\ebvii a \cr &N_3^* \phi (2v_k^{(2)})+
\sum_{j=1}^{r}{\phi(2v_k^{(2)}-2v_j^{(1)})}
-\sum_{l=1}^{s}{\phi(v_k^{(2)}-v_l^{(2)}, \gamma)} =2 \pi I_k^{(2)} ,
\ 1\leq k\leq s,	&\ebvii b \cr
}
$$}
where $I_k^{(1)}$ and $I_k^{(2)}$ are half-integers.

In the thermodynamic limit $N\rightarrow \infty$, the roots tend to
have continuous distributions. Unlike what happens in other cases, we
cannot distinguish between the roots coming from the different types
of representations, this is noted by simple inspection of the
equations of the ansatz. Due to that, we define two root densities,
one by each level,
\eqn\ebviii{
\rho_l(v_{j}^{(l)})=\lim_{N_3\rightarrow \infty}{{1 \over
{N_3(v_{j+1}^{(l)}-v_{j}^{(l)})}}}, \qquad l=1,2 \, . }

Let it be
\eqna\ebix{
$$\eqalignno{
&Z_{N_3}(v)={1 \over {2 \pi}} \Bigl[
\phi (2v)-
{1 \over N_3} \sum_{j=1}^{r}{\phi(v-v_j^{(1)})}+ {1 \over N_3}
\sum_{j=1}^{s}{\phi(2v-2v_j^{(2)})} \Bigr],
&\ebix a \cr
&Z_{N_3^*}(v)={1 \over {2 \pi}} \Bigl[
\phi (2v,)-
{1 \over N_3^*} \sum_{j=1}^{s}{\phi(v-v_j^{(2)})}+ {1 \over N_3^*}
\sum_{j=1}^{r}{\phi(2v-2v_j^{(1)})} \Bigr].
&\ebix b \cr
}$$}
In the thermodynamic limit, the derivative of these functions are
\eqna\ebx{
$$\eqalignno{
&\sigma_{1}(v)\equiv{d \over dv}{Z_{N_3}(v)}\approx {N \over
{N_3}}\rho_1(v) + {1 \over N_{3}}
\sum_{h=1}^{N_{h}^{(1)}}\delta(v-\theta_{h}^{1}), &\ebx a \cr
&\sigma_{2}(v)\equiv{d \over dv}{Z_{N_3^*}(v)}={ N \over
N_3^*}\rho_2(v) + {1 \over N_{3}^{*}}
\sum_{h=1}^{N_{h}^{(2)}}\delta(v-\theta_{h}^{2}). &\ebx b \cr
}$$}
Using the approximation
\eqn\ebxi{
\lim_{N \rightarrow \infty}{ {1 \over N} \sum_{j} f(v_j^{(k)}) \simeq
\int_{}^{}
{d\lambda f(\lambda) \rho_k(\lambda)}
},
}
and doing the Fourier transform, we can solve the system of
equations. Thus, we write
\eqna\ebxii{
$$\eqalignno{
& \sigma_{1}(v) = \sigma_{1}^{(o)}(v) + {1 \over N_{3}}
\sigma_{1}^{(h)}(v) &\ebxii a \cr
& \sigma_{2}(v) = \sigma_{2}^{(o)}(v) + {1 \over N_{3}^{*}}
\sigma_{2}^{(h)}(v), &\ebxii b \cr
}$$}
where $\sigma_{k}^{(o)}(v)$ and $\sigma_{k}^{(h)}(v)$ show the root
contribution and hole contribution respectively for $k$-level. One
finds \eqna\ebxiii
$$\eqalignno{
& \sigma_{1}^{(o)}(v) = {1 \over 2\pi} \int_{-\infty}^{+\infty}
\left({\sinh \alpha \over \sinh ({3 \alpha \over 2})}+ {N_{3}^{*}
\over N_{3}}{\sinh ({\alpha \over 2}) \over \sinh ({3 \alpha \over
2})}
\right) e^{i\alpha v} d\alpha &\ebxiii a\cr & \sigma_{2}^{(o)}(v) =
{1 \over 2\pi} \int_{-\infty}^{+\infty} \left({\sinh \alpha \over
\sinh ({3 \alpha \over 2})}+ {N_{3} \over N_{3}^{*}}{\sinh ({\alpha
\over 2}) \over \sinh ({3 \alpha \over 2})}
\right) e^{i\alpha v} d\alpha & \ebxiii b\cr & \sigma_{1}^{(h)}(v) =
{1 \over 2\pi} \left\{ \sum_{h=1}^{N_{h}^{(1)}}
r_{a}(v-\theta_{h}^{(1)}) - \sum_{h=1}^{N_{h}^{(2)}}
r_{b}(v-\theta_{h}^{(2)}) \right\} &\ebxiii c\cr
& \sigma_{2}^{(h)}(v) = {1 \over 2\pi} \left\{
\sum_{h=1}^{N_{h}^{(2)}} r_{a}(v-\theta_{h}^{(2)}) -
\sum_{h=1}^{N_{h}^{(1)}} r_{b}(v-\theta_{h}^{(1)}) \right\} ,
&\ebxiii d\cr
}
$$
with
\eqna\ebxiv
$$\eqalignno{
& r_{a}(x) = \int_{-\infty}^{+\infty}
{\sinh ({\alpha \over 2}) \over \sinh ({3 \alpha \over 2})}
e^{i\alpha x -|\alpha|} d\alpha &\ebxiv a\cr & r_{b}(x) =
\int_{-\infty}^{+\infty}
{\sinh ({\alpha \over 2}) \over \sinh ({3 \alpha \over 2})}
e^{i\alpha x +{|\alpha| \over 2}} d\alpha . & \ebxiv b\cr }
$$

\newsec{Ground state and excitations}
Any physical state is characterized by two sets of roots satisfying
the Bethe equations \ebv{a,b}. These roots can be complex and besides
we can find some modifications of the distribution of roots (holes).
So, we write the solutions of Bethe equations as strings \eqna\eci
$$\eqalignno{
& v_{k,(m)}^{(1)} = v_{k,M}^{(1)} +im \qquad ;\qquad m=-M,\ldots +M
&\eci a\cr & v_{k,(m)}^{(2)} = v_{k,M'}^{(2)} +im \qquad ;\qquad
m=-M',\ldots +M'. & \eci b\cr }
$$
A $M$-string has $M$ length and contains $M$ roots, which share the
same real part. The $0$-strings are real numbers. In order to find
the equations for the center of strings we introduce \eci{a,b}\ into
\ebv{a,b}\ and multiply the Bethe equations for all the roots of the
same string. Using the appendix A, we get \eqna\ecii
$$\eqalignno{
2N_{3}\arctan {v_{k,M}^{(1)} \over {M+ {1 \over 2}}} = & 2\pi
Q_{k,M}^{(1)} +\sum_{M'}^{} \sum_{j=1}^{\nu_{M'}^{(1)}}
\psi_{M,M'}(v_{k,M}^{(1)}-v_{j,M'}^{(1)}) & \cr & - \sum_{M''}^{}
\sum_{l=1}^{\nu_{M''}^{(2)}}
\phi_{M,M''}(v_{k,M}^{(1)}-v_{l,M''}^{(2)}) , &\ecii a \cr
-2N_{3}^{*}\arctan {v_{k,M}^{(2)} \over {M+ {1 \over 2}}} = & -2\pi
Q_{k,M}^{(2)} - \sum_{M'}^{} \sum_{j=1}^{\nu_{M'}^{(2)}}
\psi_{M,M'}(v_{k,M}^{(2)}-v_{j,M'}^{(2)}) & \cr & + \sum_{M''}^{}
\sum_{l=1}^{\nu_{M''}^{(1)}}
\phi_{M,M''}(v_{k,M}^{(2)}-v_{l,M''}^{(1)}) \, , &\ecii b \cr }
$$
where $\nu_{M}^{(i)}$ is the number of $M$-strings at level $i$, and
the numbers $Q_{k,M}^{(1)}$ and $Q_{k,M}^{(2)}$ are integers or
half-odd. They vary in the intervals $|Q_{k,M}^{(1)}| \leq
Q_{max,M}^{(1)}$ and $|Q_{k,M}^{(2)}| \leq Q_{max,M}^{(2)}$. In order
to obtain $Q_{max,M}^{(1)}$ and $Q_{max,M}^{(2)}$, we define the
functions
\eqna\eciii
$$\eqalignno{
F_{M}^{(1)}(\lambda) = &
{N_{3} \over \pi} \arctan {\lambda \over {M+ {1 \over 2}}} -{1 \over
2\pi}\sum_{M'}^{} \sum_{j=1}^{\nu_{M'}^{(1)}}
\psi_{M,M'}(\lambda-v_{j,M'}^{(1)}) & \cr & + {1 \over 2\pi}
\sum_{M''}^{} \sum_{l=1}^{\nu_{M''}^{(2)}}
\phi_{M,M''}(\lambda-v_{l,M''}^{(2)}) , &\eciii a \cr
F_{M}^{(2)}(\lambda) = &
{N_{3}^{*} \over \pi} \arctan {\lambda \over {M+ {1 \over 2}}} -{1
\over 2\pi}\sum_{M'}^{} \sum_{j=1}^{\nu_{M'}^{(2)}}
\psi_{M,M'}(\lambda-v_{j,M'}^{(2)}) & \cr & + {1 \over 2\pi}
\sum_{M''}^{} \sum_{l=1}^{\nu_{M''}^{(1)}}
\phi_{M,M''}(\lambda-v_{l,M''}^{(1)}) \,. &\eciii b \cr }
$$
The equations \ecii{a,b}\ can be written as follows \eqna\eciv
$$\eqalignno{
& F_{M}^{(1)}(v_{j,M}^{(1)}) = Q_{k,M}^{(1)} &\eciv a \cr &
F_{M}^{(2)}(v_{j,M}^{(2)}) = Q_{k,M}^{(2)} . &\eciv b \cr }
$$
Note that $F_{M}^{(1)}(\lambda)$ and $F_{M}^{(2)}(\lambda)$ are
increasing functions of $\lambda$, so we deduce \eqna\ecv
$$\eqalignno{
& F_{M}^{(1)}(-\infty) \leq Q_{k,M}^{(1)} \leq F_{M}^{(1)}(+\infty)
&\ecv a \cr & F_{M}^{(2)}(-\infty) \leq Q_{k,M}^{(2)} \leq
F_{M}^{(2)}(+\infty). &\ecv b \cr }
$$
The total number of allowed $Q_{k,M}^{(i)}$ will be %
\eqn\ecvi{2 Q_{max,M}^{(i)} + 1 = 2F_{M}^{(i)}(+\infty).} %
If we denote by $H_{M}^{(i)}$ the number of holes in the sea of
$M$-strings at level $i$, then we have
\eqn\ecvii{2 Q_{max,M}^{(i)} + 1 = \nu_{M}^{(i)}+ H_{M}^{(i)},} %
because the total number of allowed $Q_{k,M}^{(i)}$ corresponds to
the sum of roots and holes.

Taking the limit when $\lambda$ tends to infinity in \eciii{a,b}\ and
using \ecvi\ and \ecvii\ we get
\eqna\ecviii
$$\eqalignno{
& \nu_{M}^{(1)}+ H_{M}^{(1)} = N_{3} - 2\sum_{M'\geq 0}^{} J(M,M')
\nu_{M'}^{(1)} + 2\sum_{M''\geq 0}^{} K(M,M'') \nu_{M''}^{(2)}
&\ecviii a \cr
& \nu_{M}^{(2)}+ H_{M}^{(2)} = N_{3}^{*} - 2\sum_{M'\geq 0}^{}
J(M,M') \nu_{M'}^{(2)} + 2\sum_{M''\geq 0}^{} K(M,M'')
\nu_{M''}^{(1)}, &\ecviii b \cr
}
$$
with
\eqna\ecix
$$\eqalignno{
& J(M_{1},M_{2}) =\left\{ \matrix{
2M_{1}+{1 \over 2} & if & M_{1} = M_{2} \cr 2\min (M_{1},M_{2})+1 &
if & M_{1}\neq M_{2} \cr} \right. &\ecix a \cr & K(M_{1},M_{2})
=\left\{ \matrix{
M_{2}+{1 \over 2} & if & M_{2}+{1 \over 2} \leq M_{1} \cr M_{1}+{1
\over 2} & if & M_{2}+{1 \over 2} > M_{1} \cr} \right. .
&\ecix a \cr
}
$$
If $N_{\rho}$ is the number of states $\rho$ in the chain, then, as
we have shown, in a recent paper,
\ref\rxxv{J. Abad and M. R\'{\i}os, Univ. de Zaragoza preprint, DFTUZ
95-24 (1995), to appear in J. Phys A.},
\eqna\ecx
$$\eqalignno{
& N_{u}-N_{\bar{u}} = N_{3}-r &\ecx a \cr & N_{d}-N_{\bar{d}} = r-s
&\ecx b \cr
& N_{s}-N_{\bar{s}} = s-N_{3}^{*}. &\ecx c \cr }
$$
On the other hand, the total number of strings is $r$ and $s$ at
first and second level respectively, that is
\eqna\ecxi
$$\eqalignno{
& r=\sum_{M \geq 0}^{} (2M+1) \nu_{M}^{(1)} &\ecxi a \cr & s=\sum_{M
\geq 0}^{} (2M+1) \nu_{M}^{(2)}, &\ecxi b \cr }
$$
where $M$ is integer or half-odd.
Applying \ecviii{a,b}\ for the real roots, and using \ecix{a,b}\ , we
get \eqna\ecxii
$$\eqalignno{
& H_{0}^{(1)} = N_{3} - 2\sum_{M'\geq 0}^{} \nu_{M'}^{(1)} +
\sum_{M''\geq 0}^{} \nu_{M''}^{(2)} &\ecxii a \cr & H_{0}^{(2)} =
N_{3}^{*} - 2\sum_{M'\geq 0}^{} \nu_{M'}^{(2)} + \sum_{M''\geq 0}^{}
\nu_{M''}^{(1)}. &\ecxii b \cr }
$$
For the general case this relation can be written as \eqna\ecxiii
$$\eqalignno{
& \nu_{n}^{(1)}-{\nu_{n}^{(2)} \over 2}+ H_{n}^{(1)} =
{{H_{n-1/2}^{(1)}+H_{n+1/2}^{(1)}} \over 2} &\ecxiii a \cr &
\nu_{n}^{(2)}-{\nu_{n}^{(1)} \over 2}+ H_{n}^{(2)} =
{{H_{n-1/2}^{(2)}+H_{n+1/2}^{(2)}} \over 2} &\ecxiii b \cr & \qquad
\qquad n=0,{1 \over 2},1,{3 \over 2},2,\ldots, & \cr }
$$
where we have used $H_{-1/2}^{(1)}\equiv N_{3}$ and
$H_{-1/2}^{(2)}\equiv N_{3}^{*}$.

At this moment, we can characterize the ground state and the excited
states:

Firstly, {\it Ground state}. In the ground state we have no holes and
only real roots. That is
\eqna\ecxiv
$$\eqalignno{
& \nu_{0}^{(1)} = {2N_{3}+N_{3}^{*} \over 3} &\ecxiv a \cr &
\nu_{0}^{(2)} = {N_{3}+2N_{3}^{*} \over 3} &\ecxiv b \cr &
\nu_{M>0}^{(1)}= \nu_{M>0}^{(2)} =H_{M \geq 0}^{(1)} = H_{M \geq
0}^{(2)} =0, &\ecxiv c \cr
}
$$
and the quantum numbers are
\eqn\ecxv{N_{u}-N_{\bar{u}}=N_{d}-N_{\bar{d}}=N_{s}-N_{\bar{s}}=
{N_{3}-N_{3}^{*} \over 3}.}
Then, the ground state is formed by pairs $u\bar{u}$, $d\bar{d}$ and
$s\bar{s}$.

Secondly, {\it Excited state with the same quantum numbers}. This is
characterized by one hole and one two-string in each level,
\eqna\ecxvi
$$\eqalignno{
& \nu_{0}^{(1)} = {2N_{3}+N_{3}^{*} \over 3}-2 &\ecxvi a \cr &
\nu_{0}^{(2)} = {N_{3}+2N_{3}^{*} \over 3}-2 &\ecxvi b \cr &
\nu_{1/2}^{(1)}= \nu_{1/2}^{(2)} =H_{0}^{(1)} = H_{0}^{(2)} =1,
&\ecxvi c \cr
& \nu_{M>1/2}^{(1)}= \nu_{M>1/2}^{(2)} =H_{M > 0}^{(1)} = H_{M >
0}^{(2)} =0, &\ecxvi d \cr
}
$$
and the quantum numbers as in \ecxv\ .

Thrirdly, {\it Excited state with other quantum numbers}. We can find
an excited state characterized by two holes, one in each level, and
real number, that is
\eqna\ecxvii
$$\eqalignno{
& \nu_{0}^{(1)} = {2N_{3}+N_{3}^{*} \over 3}-1 &\ecxvii a \cr &
\nu_{0}^{(2)} = {N_{3}+2N_{3}^{*} \over 3}-1 &\ecxvii b \cr &
H_{0}^{(1)} = H_{0}^{(2)} =1, &\ecxvii c \cr & \nu_{M>0}^{(1)}=
\nu_{M>0}^{(2)} =H_{M > 0}^{(1)} = H_{M > 0}^{(2)} =0. &\ecxvii d \cr
}
$$
Here, the quantum numbers are
\eqna\ecxviii
$$\eqalignno{
& N_{u}-N_{\bar{u}}= {N_{3}-N_{3}^{*} \over 3}+1 &\ecxviii a \cr &
N_{d}-N_{\bar{d}}= {N_{3}-N_{3}^{*} \over 3} &\ecxviii b \cr &
N_{s}-N_{\bar{s}}= {N_{3}-N_{3}^{*} \over 3}-1. &\ecxviii c \cr }
$$
There are four ways to get this state from the ground state. The
first is when a $d$-site changes to an $u$-site and a $\bar{d}$-site
to a $\bar{s}$-site. The second when it changes from an $\bar{u}$ to
a $\bar{d}$ and a $s$ to $d$. The third from a $s$ to $u$, and the
fourth is a $\bar{u}$ to a $\bar{s}$.

The next excited states have more that two holes, and they can be
found by using \ecxi\ to \ecxiii\ .

\newsec{S-matrix}
In this section we are going to find the S-matrix for the excitations
over the ground state. For this purpose we will use the transfer
matrix $\bar{F}(u,\gamma)$, which is obtained by taking the
$\{3^{*}\}$-representation in the auxiliary space. Thus, we define %
\eqn\edi{ \bar{T}_{\alpha,\beta}(u,\gamma) =
L_{\alpha,\alpha_{1}}^{(\{3^{*}\},\{3\})}(u+\gamma)
L_{\alpha_{1},\alpha_{2}}^{(\{3^{*}\},\{3^{*}\})}(u) \ldots
L_{\alpha_{N-2},\alpha_{N-1}}^{(\{3^{*}\},\{3\})}(u+\gamma)
L_{\alpha_{N-1},\beta}^{(\{3^{*}\},\{3^{*}\})}(u), }
and the transfer matrix is
\eqn\edii{\bar{F}(u,\gamma) = \trace [\bar{T}(u,\gamma)]. }
>From the Yang-Baxter equation we have that
\eqn\ediii{[T(u,\gamma),\bar{T}(v,-\gamma)]=0. }
We take the simplest case, that is $\gamma=0$.

The eigenvalue of $\bar{F}(u,0)$ is
\eqnn\ediv
$$\eqalignno{\bar{\Lambda}(u) =
& [a(u)]^{N_{3}^{*}}
	 [\bar{b}(u)]^{N_{3}}
	 \prod_{j=1}^{\bar{r}} g(\bar{\mu}_{j}-u) + [b(u)]^{N_{3}^{*}}
\prod_{i=1}^{\bar{r}} g(u-\bar{\mu}_{i}) \times & \cr & \left\{
	 [\bar{b}(u)]^{N_{3}} \prod_{l=1}^{\bar{s}} g(\bar{\lambda}_{l}-u)
+ 	 [\bar{a}(u)]^{N_{3}} \prod_{k=1}^{\bar{s}}
g(u-\bar{\lambda}_{k}) 	 \prod_{n=1}^{\bar{r}} {1 \over
g(u-\bar{\mu}_{n})} 	 \right\} , \qquad & \ediv \cr
}
$$
and, using the parameterization \ebiv\ , the ansatz equations can be
written as
\eqna\edv
$$\eqalignno{
& \left[{{\bar{v}_{k}^{(1)}-{i \over 2}} \over {\bar{v}_{k}^{(1)}+{i
\over 2}}}\right]^{N_{3}^{*}} =- \prod_{j=1}^{\bar{r}}
{{\bar{v}_{k}^{(1)}- \bar{v}_{j}^{(1)}-i}
\over {\bar{v}_{k}^{(1)}-\bar{v}_{j}^{(1)}+i}} \prod_{l=1}^{\bar{s}}
{{\bar{v}_{l}^{(2)}-\bar{v}_{k}^{(1)}-{i \over 2}} \over
{\bar{v}_{l}^{(2)}-\bar{v}_{k}^{(1)}+{i \over 2}}} , &\edv a\cr
& \left[{{\bar{v}_{k}^{(2)}+{i \over 2}} \over {\bar{v}_{k}^{(2)}-{i
\over 2}}}\right]^{N_{3}} =- \prod_{j=1}^{\bar{r}}
{{\bar{v}_{k}^{(2)}- \bar{v}_{j}^{(1)}-{i \over 2}}
\over {\bar{v}_{k}^{(2)}-\bar{v}_{j}^{(1)}+{i \over 2}}}
\prod_{l=1}^{\bar{s}} {{\bar{v}_{l}^{(2)}-\bar{v}_{k}^{(2)}-i} \over
{\bar{v}_{l}^{(2)}-\bar{v}_{k}^{(2)}+i}} . & \edv b\cr }
$$
But the commutation of the transfer matrices in \ediii\ requires that
the Bethe equations \ebv\ and \edv\ are the same. Thus, we have
\eqna\edvi
$$\eqalignno{
& \bar{v}_{j}^{(1)} = v_{j}^{(2)} \quad j=1,\ldots,\bar{r} = s ,
&\edvi a\cr & \bar{v}_{k}^{(2)} = v_{k}^{(1)} \quad
k=1,\ldots,\bar{s} = r . & \edvi b\cr }
$$

In order to calculate the momentum of the chain, we consider an
alternating chain, that is $N_{3}=N_{3}^{*}= {N \over 2}$. Then, the
momentum is
\eqn\edvii{P= i \ln \left[ \bar{\rho}(0)^{-N/2} \Lambda (0)
\bar{\Lambda}(0)
\right].
}
With \ebi , \ebiv{a,b}, \ediv\ and \edvi{a,b}\ we have %
\eqn\edviii{P= i \sum_{j=1}^{r} \ln {{v_{j}^{(1)}-{i \over 2}} \over
{v_{j}^{(1)}+{i \over 2}}} + i \sum_{k=1}^{s} \ln {{v_{k}^{(2)}- {i
\over 2}} \over {v_{k}^{(2)}+{i \over 2}}} . }
Using the approximation \ebxi\ and the root densities \ebxiii\ , we
get
\eqn\edix{P= P_{0} + \sum_{h=1}^{N_{h}^{(1)}} p(\theta_{h}^{(1)}) +
\sum_{h=1}^{N_{h}^{(2)}} p(\theta_{h}^{(2)}), }
where $P_{0}$ is the momentum of the ground state and the other terms
are the hole contributions, with
\eqn\edx{p(\theta) = {1 \over 2} \int_{-\infty}^{+\infty} dx {{\sinh
x +\sinh ({x \over 2})} \over {ix \sinh ({3x \over 2})}} e^{ix\theta}.
}

We calculate the $S$ matrix for the scattering of two holes. Here we
follow the Korepin-Andrei-Destri method
\ref\rxviii{V.E. Korepin, Theor. Math. Phys. 41, (1979) 953.}-
\nref\rxix{V.E. Korepin, G. Izergin and N.M. Bogoliubov, {\it Quantum
Inverse Scattering Method and Correlation Functions }, Cambridge
Univ. Press, 1993.} \ref\rxx{N. Andrei and C. Destri, Nucl. Phys. B
231, (1984) 445.}. For a state of two holes
with rapidities $\theta_{1}$ and $\theta_{2}$, the momentum
$p(\theta_{1})$ verifies the quantization condition
\eqn\edxi{e^{ip(\theta_{1})N}S = 1 \,.
}
The $S$ matrix can be written as $S=e^{i\Phi}$, so we have
\eqn\edxii{p(\theta_{1}) + {1 \over N} \Phi = {2\pi \over N} n, }
where n is an integer.

One can prove by direct calculation that
\eqn\edxiii{p(\theta) = \pi \int_{-\infty}^{\theta}
\sigma_{1}^{(o)}(\lambda) d\lambda + c_{1} \,, }
where $c_{1}$ is a constant, which will be irrelevant for our
problem.
>From \ebx{a}\ we can write
\eqn\edxiv{Z_{N/2}(\theta) = \int_{-\infty}^{\theta}
\sigma_{1}(\lambda) d\lambda +c_{2},
}
where $c_{2}$ is an irrelevant constant.

Evaluating for a hole in the first level $\theta_{1}$ we have
\eqn\edxv{Z_{N/2}(\theta_{1}) = {I^{(h)} \over N/2}. }
With the relations \edxiii\ , \edxiv\ and \edxv\ we deduce
\eqn\edxvi{p(\theta_{1}) = {2\pi \over N} I^{(h)} - {2\pi \over N}
\int_{-\infty}^{\theta_{1}} \sigma_{1}^{(h)}(\lambda) d\lambda +
const. }
Comparing \edxii\ with \edxvi\ , we conclude
\eqn\edxvii{\Phi = 2 \pi \int_{-\infty}^{\theta_{1}}
\sigma_{1}^{(h)}(\lambda) d\lambda + const. }
We remove the constants because they contribute as a
rapidity-independent phase factor. Thus we have to calculate
\eqn\edxviii{\Phi(\theta) = -2 \int_{0}^{\infty} {d\alpha \over
\alpha} {\sinh {\alpha \over 2} \over \sinh {3\alpha \over 2}}
e^{\alpha \over 2} \sin (\alpha \theta), }
where $\theta=\theta_{1}-\theta_{2}$ is the difference of rapidities
of the two holes. This integral can be solved by means of the
$\Psi$-function
\eqn\edxix{\Psi (x) = {d \over dx} \ln \Gamma (x), }
and we get
\eqn\edxx{{d\Phi \over d\theta} = {1 \over 3} \left[ -\Psi({1 \over
6}-i {\theta \over 3})+
\Psi({1 \over 2}-i {\theta \over 3}) -
\Psi({1 \over 6}+i {\theta \over 3})+
\Psi({1 \over 2}+i {\theta \over 3})
\right]\, .
}
For the corresponding $S$ matrix we have %
\eqn\edxxi{S_{1}(\theta) = {{\Gamma({1 \over 6}-i {\theta \over 3})
\Gamma({1 \over 2}+i {\theta \over 3})}
\over {\Gamma({1 \over 6}+i {\theta \over 3})} \Gamma({1 \over 2}-i
{\theta \over 3})} . }
For the state with one hole and one two-string in each level we find
the scattering  matrix to be
\eqn\edxxii{S_{2}(\theta) = {({1 \over 2}-i\theta) \over
({1 \over 2}+i\theta)}
{{\Gamma({1 \over 6}-i {\theta \over 3})
\Gamma({1 \over 2}+i {\theta \over 3})}
\over {\Gamma({1 \over 6}+i {\theta \over 3})} \Gamma({1 \over 2}-i
{\theta \over 3})} . }
In order to calculate the $S$ matrix for holes in the same level we
consider states with, at least, four holes (two in each level).
Following the same procedure  we find for two holes in the first
(second) level

\eqn\edxxi{S_{3}(\theta) = {{\Gamma({2 \over 3}-i {\theta \over 3})
\Gamma(1+i {\theta \over 3})}
\over {\Gamma({2 \over 3}+i {\theta \over 3})} \Gamma(1-i
{\theta \over 3})} , }
where $\theta=\theta_{1}^{(1)}-\theta_{2}^{(1)}$ ($\theta=
\theta_{1}^{(2)}-\theta_{2}^{(2)}$).
The $S_{1}$ and $S_{3}$ matrices coincide with those for the no-alternating
chain.
This shows that the scattering is the same in the alternating and the
non alternating chain \rvi.
\bigbreak\bigskip\bigskip{{\bf Acknowledgments}}\nobreak

This work was partially supported by the Direcci\'on General de
Investigaci\'on Cient\'{\i}fica y T\'ecnica, Grant No PB93-0302.

\appendix{A}{}
We define the function
\eqn\epi{ V_{0}(x) = {x-i \over x+i}.}
It is easy to prove the relations
\eqna\epii
$$\eqalignno{
& \prod_{m=-M}^{M} V_{0}(2x+i2m) =V_{0}({x \over M+{1 \over 2}})
&\epii a \cr & \prod_{m=-M}^{M} V_{0}(x+im) =V_{0}({x \over M})
V_{0}({x \over M+1}) \equiv V_{M}(x) &\epii b \cr &
\prod_{m_{1}=-M_{1}}^{M_{1}} \prod_{m_{2}=-M_{2}}^{M_{2}}
V_{0}(x+i(m_{1}+m_{2})) =\prod_{L=|M_{2}-M_{1}|}^{M_{1}+M_{2}}
V_{L}(x) \equiv V_{M_{1},M_{2}}(x) &\epii c \cr &
\prod_{m_{1}=-M_{1}}^{M_{1}} \prod_{m_{2}=-M_{2}}^{M_{2}}
V_{0}(2x+i2(m_{1}+m_{2})) =\prod_{m_{1}=-M_{1}}^{M_{1}}
V_{0}({x+im_{1} \over M_{2}+{1 \over 2}}) \equiv W_{M_{1},M_{2}}(x).
\qquad &\epii d \cr
}
$$
It is convenient to define the functions \eqna\epiii
$$\eqalignno{
& \psi_{M_{1},M_{2}}(x) = 2\sum_{L=|M_{2}-M_{1}|}^{M_{1}+M_{2}}
\left( \arctan {x \over L} + \arctan {x \over L+1} \right) &\epiii a
\cr & \phi_{M_{1},M_{2}}(x) = 2\sum_{L=-M_{1}}^{M_{1}}
\arctan ({x+iL \over M_{2}+{1 \over 2}}), &\epiii b \cr }
$$
which are connected with \epii{c,d}\ by the logarithm.

\listrefs
\end{document}